\documentclass{article} 
\usepackage[final]{colm2026_conference}

\usepackage{microtype}
\usepackage{hyperref}
\usepackage{url}
\usepackage{booktabs}
\usepackage{color,soul}
\usepackage{xcolor}
\usepackage[table]{xcolor}
\usepackage{graphicx}
\definecolor{lightgreen}{RGB}{201, 223, 189} 
\usepackage{longtable}
\usepackage{array}
\usepackage{arydshln}
\usepackage[most]{tcolorbox}
\usepackage{enumitem}
\usepackage{microtype}
\usepackage{placeins}
\usepackage{multirow}
\usepackage{verbatim}
\usepackage{float}
\usepackage{lineno}

\definecolor{darkblue}{rgb}{0, 0, 0.5}
\hypersetup{colorlinks=true, citecolor=darkblue, linkcolor=darkblue, urlcolor=darkblue}

\title{Evaluating the Impact of Personalization in Conversational Cybersecurity Assistants}

\author{Lea Duesterwald\thanks{Home institution: Cornell University.}\\
Carnegie Mellon University\\
\texttt{lduester@andrew.cmu.edu}
\And
Anika Jain\\
Carnegie Mellon University\\
\texttt{anikajai@andrew.cmu.edu}
\And
Shreya Kochar\phantom{XXXXXXXXXXXXXXXXXXXXXXXXXXlll}\\
Carnegie Mellon University\\
\texttt{shreyako@andrew.cmu.edu}
\And
Norman Sadeh\\
Carnegie Mellon University\\
\texttt{sadeh@cs.cmu.edu}
}

\begin{document}

\ifcolmsubmission
\linenumbers
\fi

\maketitle

\title{Evaluating the Impact of Personalization in Conversational Cybersecurity Assistants}


\begin{abstract}
  Users increasingly turn to Large Language Models to answer a variety of questions, including cybersecurity questions. We study how personalization strategies can help improve the effectiveness of answers to questions asked to an LLM-based cybersecurity assistant. Beyond accuracy, we focus on the understandability, actionability and, most importantly, motivating power of answers, given how often users fail to follow cybersecurity recommendations. Specifically, we investigate four personalization strategies, ranging from static user profiles to interaction-history-based personalization, using a corpus of 1,045 real-world cybersecurity questions and a 7-day deployment involving 57 participants and 1,066 user questions. Across both a large-scale automated LLM-based evaluation and human evaluation, conversation-based personalization is consistently favored in comparative ratings of perceived helpfulness and likelihood of following security advice. Importantly, the relative trends observed in the LLM-based evaluation align with those obtained from human evaluation, suggesting that LLM-based evaluation can provide a scalable mechanism for comparing personalization strategies before costly user studies. These results indicate that behavior-driven personalization is a promising direction for LLM-powered cybersecurity assistants and highlight the value of combining LLM-based and human evaluation when studying personalized language-model systems.
\end{abstract}

\section{Introduction}

Large language models are increasingly used as assistants for cybersecurity questions, creating opportunities to provide users with timely and actionable guidance. Prior work has shown that prompt engineering grounded in Protection Motivation Theory (PMT) can improve the motivational quality of LLM-generated security advice and increase users' self-reported likelihood of following recommendations \citep{duesterwald2025}. However, even with PMT-based prompting, many users still fail to act on the advice they receive. Cybersecurity often remains a secondary task to users' main online activities \citep{convenience}, highlighting the need for approaches that not only provide accurate information, but also increase users' willingness to act on it.

Personalization offers a potential mechanism for improving the effectiveness of LLM-powered cybersecurity assistants. Existing approaches range from static user profiles to methods that adapt based on accumulated interaction history \citep{chen2024,zhang2025memory,zhao2025teaching}. We investigate how different forms of personalization affect LLM-powered cybersecurity assistants and whether trends observed through LLM-based evaluation correspond to evaluations by real users.

In this work, we investigate four personalization strategies for LLM-based cybersecurity assistants, including profile-based approaches, expertise-aware few-shot prompting, and conversation-based personalization that conditions on prior questions, answers, and user feedback. We first conduct a large-scale automated LLM-based evaluation using a corpus of 1,045 real-world cybersecurity questions collected in our prior work \citep{duesterwald2025}. LLM-based evaluation results identify interaction-history-based personalization as the strongest approach across understandability, actionability, and motivating power. Because LLM-based evaluations alone cannot fully capture human perceptions, we validate these findings through a 7-day deployment involving 57 participants and 1,066 questions. Across both studies, conversation-based personalization improves comparative ratings of perceived helpfulness and users' self-reported likelihood of following security advice, while static background-based personalization provides limited benefits relative to a strong PMT-based baseline.

Our contributions are threefold. First, we compare four personalization strategies for LLM-based cybersecurity assistants and identify interaction-history-based personalization as the most effective approach. Second, we show that conversation-based personalization improves comparative ratings of perceived helpfulness and users' self-reported likelihood of following security advice, outperforming both a PMT-only baseline and static background-based personalization. Third, we demonstrate that trends observed through large-scale automated LLM-based evaluations align with those obtained from human evaluation, suggesting that LLM-based evaluation can provide a scalable mechanism for comparing personalization strategies before costly user studies.

Together, these results suggest that behavior-driven personalization is a promising direction for LLM-powered cybersecurity assistants and highlight the value of combining LLM and human evaluation when studying personalized language-model systems.

\section{Review of the Literature}
\label{sec:review_of_lit}

Users frequently deprioritize security because it competes with primary goals such as productivity and task completion \citep{acquisti2015}. As a result, providing accurate information alone is often insufficient to influence behavior, creating a gap between knowledge and action \citep{redmiles2016census}. Prior work has shown that conversational systems can provide contextually relevant cybersecurity guidance \citep{redmiles2016advice}, and recent work demonstrated that prompt engineering based on PMT can improve the motivational quality of responses and increase users' self-reported likelihood of following security advice \citep{duesterwald2025}. These findings suggest that language models can serve not only as information sources but also as mechanisms for encouraging protective behavior.

Personalization offers a potential way to further improve the effectiveness of LLM-generated advice. Prior work has shown that personalized interaction strategies increase user engagement and perceived usefulness in interactive systems \citep{knijnenburg2012,kaptelinin2015}. In language models, personalization can be achieved using explicit user information, few-shot examples, or accumulated interaction histories \citep{brown2020,xu2026personalized}. Memory-based approaches enable systems to adapt over time by leveraging previous interactions and user feedback, supporting more persistent and behavior-driven forms of personalization \citep{chen2024,zhang2025memory,zhao2025teaching}. Building on prior work in personalization and adaptive systems, this work systematically compares multiple personalization strategies and examines their effects on answer quality, motivating power, perceived usefulness, and users' willingness to follow cybersecurity advice. We additionally investigate whether trends observed through LLM-based evaluations correspond to human evaluations.

\section{Combining LLM and Human-Based Evaluations of Personalization Strategies}
\label{sec:pers_description}

We adopt a two-stage evaluation framework that combines large-scale LLM-based evaluations with an in situ user study. LLMs enable efficient evaluations of multiple personalization strategies across a large corpus of questions, while the user study validates whether trends observed through automated evaluation translate to real users.

We used a publicly available corpus of 1,045 real-world cybersecurity questions and corresponding answers from our previous work for this evaluation \citep{duesterwald2025}. The dataset additionally includes user background information and behavioral responses, enabling the construction of several forms of personalization. For each question, we generated new answers using different personalization strategies and evaluated them using an automated scoring framework.

\subsection{Personalization Strategies}

We examined four personalization approaches which span both static and dynamic sources of personalization. Complete prompts and contexts are provided in Appendix~\ref{sec:full_context_pers_appendix}.

\begin{itemize}
\item \textbf{P1: Background-Based Personalization (6 Questions).} P1 augments a strong Protection Motivation Theory (PMT)-based prompt developed in prior work \citep{balaji2024} with six intake survey responses selected through principal component analysis. The LLM uses these responses to adapt its answers to users' self-reported characteristics, beliefs, and behaviors.

\item \textbf{P2: Background-Based Personalization (56 Questions).} P2 extends P1 by incorporating the complete set of 56 intake survey responses.

\item \textbf{P3: Knowledge-Level Few-Shot Personalization.} Users are categorized into low, medium, or high technical knowledge groups based on self-reported familiarity with common security concepts. Tailored example responses are then included in the prompt to encourage explanations appropriate for each expertise level.

\item \textbf{P4: Conversation-Based Personalization.} P4 performs dynamic personalization based on accumulated interaction history. For each user, the context includes prior questions, generated answers, alternate answers, and user evaluations, allowing the system to adapt using behavioral signals observed during interaction.

\end{itemize}

\subsection{LLM-Based Evaluation Procedure}

For each personalization strategy, we generated answers to all questions and evaluated them using GPT-4-turbo. Following prior work \citep{balaji2024}, answers were assessed along three dimensions: \textit{understandability}, \textit{actionability}, and \textit{motivating power}. For each question, the evaluator compared the personalized answer against the PMT-only baseline without knowledge of their origin. For selected strategies, we additionally obtained independent ratings. Full evaluation prompts and rating scales are provided in Appendix~\ref{sec:eval_prompts_appendix}. We also compare the direction and relative magnitude of the two personalization contrasts represented in both the LLM-based and human evaluations; this cross-method comparison is reported in Section~\ref{sec:results_main} and Appendix~\ref{sec:cross_method_consistency_appendix}.

\section{Security-Based Question Answering Assistant Design}

We implemented a browser-based cybersecurity assistant that enables users to ask security-related questions and receive responses generated by GPT-4.1. Submitted questions were first checked for relevance to computer security and then processed using PMT-based prompting \citep{balaji2024} together with the personalization mechanisms described in Section~\ref{sec:pers_description}. All interactions were logged for later analysis.

For answer generation, user questions and prior message history were provided to GPT-4.1 with a temperature of 0.7 and a maximum response length of 500 tokens. During the interaction phase, only message history from the current day was included in each API call. This design allowed users to ask follow-up questions while preventing unintended carryover effects between experimental conditions. Conversation-based personalization evaluated during the final survey was constructed separately from the complete 7-day interaction history (see section~\ref{sec:user_study} for more detail).

Participants interacted with the assistant through a chat-style interface embedded within the Chrome browser. The interface supported conversational question answering and was designed to encourage users to ask cybersecurity questions arising from their daily online activities.

\section{User Study and Evaluation}
\label{sec:user_study}

To validate whether trends observed in LLM-based evaluations translate to real users, we conducted a 7-day in situ study in which participants asked cybersecurity questions arising from their daily online activities and provided structured feedback on the answers they received. This study was approved by the Carnegie Mellon University Institutional Review Board (IRB study number: STUDY2024$\textunderscore$00000291). A total of 57 participants completed the study and contributed 1,066 total questions.

Participants first completed an intake survey, which provided the information used for background-based personalization. They then interacted with the browser-based assistant for seven days, asking at least two unique cybersecurity questions per day and providing feedback on the responses. During this phase, participants were assigned to either a control condition, which used PMT-based prompting without additional personalization, or an experimental condition, which incorporated background-based personalization (P1). These interactions produced the histories later used for conversation-based personalization.

In addition to the quantitative evaluations, we conducted optional post-study interviews with 10 participants to provide qualitative context for the findings presented in Section~\ref{sec:results}.

\subsection{Final Survey: Evaluation of Conversation-Based Personalization}
\label{sec:final_survey}

The primary evaluation of conversation-based personalization (P4) was conducted after participants completed the first interaction phase. Each participant was presented with five questions they had previously asked, together with the original answer they had received and a newly generated answer based on their accumulated interaction history.

Participants directly compared the two answers in terms of helpfulness and likelihood of following the advice. For control-condition participants, the original answer corresponded to the PMT-only baseline, while for experimental-condition participants it corresponded to PMT plus background-based personalization (P1). In both cases, participants compared the original answer against a conversation-based personalized answer (P4).

To select the five questions for each participant used in this evaluation, GPT-5.2 was used to identify context-grounded questions that reflected participants' own online activities and requested specific recommendations, since our final-survey evaluation focuses on participants’ assessments of the advice and their likelihood of following it. For each participant, five questions were then randomly drawn from the resulting set. For each evaluated question, the personalization context consisted of all other questions asked during the study together with the corresponding answers and user evaluations, yielding histories containing at least 13 prior interactions.

Because each participant evaluated five questions in the final survey, we accounted for repeated measurements within participants in the statistical analysis. We primarily fit linear mixed-effects models with participant-level random intercepts. When a mixed-effects model produced a singular boundary fit, we instead report the corresponding participant-level analysis, which averages repeated evaluations within each participant before testing. We applied the Benjamini--Hochberg procedure across the seven final-survey hypothesis tests; participant-level analyses produced the same qualitative conclusions.

\section{Results}
\label{sec:results}

\subsection{Automated LLM-Based Evaluation Results}
\label{sec:results_pers_eval}

We first conducted a large-scale automated LLM-based evaluation of multiple personalization strategies. For this analysis, we used the corpus of 1,045 real-world cybersecurity questions from our prior work \citep{duesterwald2025}. For each question, we generated answers using different personalization approaches and evaluated them using GPT-4-turbo across three dimensions of answer effectiveness: understandability, actionability, and motivating power, following prior work.

\paragraph{Comparative Evaluation Across Personalization Strategies.}

\begin{table}[t]

\centering
\small
\renewcommand{\arraystretch}{1.15}
\setlength{\tabcolsep}{2pt}

\begin{tabular}{|p{0.04\columnwidth}|
p{0.14\columnwidth}p{0.15\columnwidth}|
p{0.12\columnwidth}p{0.15\columnwidth}|
p{0.14\columnwidth}p{0.15\columnwidth}|}
\hline
& \multicolumn{2}{c|}{\textbf{Understandability}}
& \multicolumn{2}{c|}{\textbf{Actionability}}
& \multicolumn{2}{c|}{\textbf{Motivating Power}} \\
\cline{2-7}
& \textbf{Mean} & \textbf{p-value}
& \textbf{Mean} & \textbf{p-value}
& \textbf{Mean} & \textbf{p-value} \\
\hline

\rowcolor{lightgreen}
\cellcolor{white}P1
& -0.334 & 1.07E-73
& -0.403 & 6.07E-77
& -0.644 & 1.04E-129 \\ \hline

\rowcolor{lightgreen}
\cellcolor{white}P2
& -0.059 & 3.85E-14
& -0.193 & 5.94E-22
& -0.388 & 6.66E-61 \\ \hline

\rowcolor{lightgreen}
\cellcolor{white}P3
& 0.339 & 1.10E-39
& 0.359 & 1.34E-48
& 0.481 & 2.17E-69 \\ \hline

\rowcolor{lightgreen}
\cellcolor{white}P4
& 0.547 & 3.22E-56
& 0.477 & 6.54E-46
& 0.709 & 3.89E-68 \\ \hline

\end{tabular}
\caption{LLM-based evaluation comparing personalized and non-personalized answers across three effectiveness metrics: understandability, actionability, and motivating power. GPT-4-turbo directly compared the two answers on a -2 to 2 scale, where positive values favor the personalized answer. Green cells indicate statistically significant differences from 0 ($p<0.05$).}
\label{tab:comparative_llm_based_results}
\end{table}

Table~\ref{tab:comparative_llm_based_results} presents comparative evaluations across the four personalization strategies (P1--P4), where the model directly compared personalized answers against non-personalized PMT-based answers. Across all three dimensions, clear and consistent patterns emerge.

Background-based personalization strategies (P1 and P2) yield negative mean scores across all dimensions, indicating that these approaches are, on average, rated as worse than the non-personalized baseline. In contrast, knowledge-level few-shot personalization (P3) produces positive scores, suggesting improvements over the baseline. Conversation-based personalization (P4) achieves the strongest performance, with the highest positive scores across all three dimensions. Notably, all observed differences are statistically significant.

These results establish a clear ordering of effectiveness among the evaluated strategies: P1 and P2 underperform the baseline, P3 provides moderate improvements, and P4 consistently yields the largest gains. In particular, the magnitude of improvement for P4 in motivating power suggests that conditioning on accumulated interaction history may meaningfully enhance the behavioral impact of generated responses.

\paragraph{Absolute Evaluation of Selected Strategies.}
\begin{table}[t]
\centering
\small
\renewcommand{\arraystretch}{1.0}
\setlength{\tabcolsep}{4pt}

\begin{tabular}{|l|ccc|ccc|}
\hline
& \multicolumn{3}{c|}{\textbf{P1}} & \multicolumn{3}{c|}{\textbf{P4}} \\
\cline{2-7}
\textbf{Metric}
& \textbf{Mean N.}
& \textbf{Mean P.}
& \textbf{$p$}
& \textbf{Mean N.}
& \textbf{Mean P.}
& \textbf{$p$} \\
\hline

\textbf{Understandability}
& \cellcolor{lightgreen}8.848
& \cellcolor{lightgreen}8.372
& \cellcolor{lightgreen}2.55E-95
& \cellcolor{lightgreen}8.618
& \cellcolor{lightgreen}8.853
& \cellcolor{lightgreen}1.56E-17
\\
\hline

\textbf{Actionability}
& \cellcolor{lightgreen}8.708
& \cellcolor{lightgreen}8.584
& \cellcolor{lightgreen}8.93E-09
& \cellcolor{lightgreen}8.425
& \cellcolor{lightgreen}8.873
& \cellcolor{lightgreen}3.03E-24
\\
\hline

\textbf{Motivating Power}
& \cellcolor{lightgreen}8.248
& \cellcolor{lightgreen}8.070
& \cellcolor{lightgreen}1.00E-21
& \cellcolor{lightgreen}8.124
& \cellcolor{lightgreen}8.687
& \cellcolor{lightgreen}5.39E-42
\\
\hline

\end{tabular}
\caption{Independent GPT-4-turbo ratings of non-personalized (Mean N.) and personalized (Mean P.) answers for background-based personalization (P1) and conversation-based personalization (P4). Answers were rated on a 1--10 scale, where higher values indicate better responses. Green cells indicate statistically significant differences ($p<0.05$).}
\label{tab:pers_eval_absolute_results}
\end{table}

To further examine these trends, we conducted an additional analysis for two representative strategies: background-based personalization (P1) and conversation-based personalization (P4). Table~\ref{tab:pers_eval_absolute_results} reports independent ratings of personalized and non-personalized answers for these strategies.

These results reinforce the patterns observed in the comparative analysis. For P1, personalized answers receive consistently lower ratings than their non-personalized counterparts across all three dimensions. In contrast, for P4, personalized answers receive significantly higher ratings than non-personalized answers, again across all dimensions.

Taken together, the comparative and absolute analyses provide consistent evidence that personalization based on accumulated interaction history is the most effective strategy among those evaluated, while static background-based approaches may not provide sufficient signal to improve response quality. Importantly, however, these findings are based on LLM-based evaluation. To determine whether these trends translate to real user perceptions and behavior, we validated these results through an in situ user study.

\subsection{In Situ Study Overview}

We next conducted a 7-day in situ user study to evaluate personalization strategies in a real-world setting and to collect the interaction history required for conversation-based personalization.

A total of 168 participants completed the initial intake survey, of whom 97 participants (49 experimental, 48 control) completed at least one day of interaction. Of these, 57 participants (35 experimental, 22 control) completed all 7 days and the final survey. Each evaluated five previously asked questions, yielding 285 question-level evaluations: 175 from the experimental group and 110 from the control group. Questions were distributed throughout the day (Appendix \ref{sec:question_time_dist_appendix}), suggesting that participants engaged with the assistant in response to real-world concerns as they arose. Across the interaction phase, participants asked 1,066 questions spanning a wide range of cybersecurity topics (Appendix \ref{sec:q_topic_dist_appendix}).

\subsection{Interaction-Phase Results}

During the interaction phase, participants evaluated answers under two conditions: PMT-only prompting (control) and background-based personalization (experimental). These results are reported in Table~\ref{tab:evening_survey_results_alt} in Appendix \ref{sec:evening_survey_res_appendix}.

Overall, results from this phase indicate that answer quality was already high in the baseline condition. Participants' evaluations did not show a strong preference in either direction: evaluations of the answer users received during the day tended to slightly favor the baseline, while direct comparisons between baseline and personalized answers tended to slightly favor background-based personalization. Consistent with the LLM-based evaluations, these findings suggest that static, background-based personalization provides mixed results when compared to an already strong PMT-based baseline.

Importantly, the primary role of the interaction phase was to collect user feedback and behavioral signals for constructing the interaction histories used in the evaluation of conversation-based personalization, which we examine next.

\subsection{Effect of Conversation-Based Personalization: Results}
\label{sec:results_main}

Conversation-based personalization (P4) yielded statistically significant improvements in comparative ratings of perceived helpfulness and self-reported likelihood of following advice against both baselines (Table~\ref{tab:final_survey_results_p4}).

In the final survey, participants compared answers generated using accumulated interaction history against the answers they originally received during the study. This enabled two comparisons: (1) no personalization versus conversation-based personalization (control group), and (2) background-based personalization versus conversation-based personalization (experimental group). Across both groups, conversation-based personalization consistently improved comparative ratings of helpfulness and likelihood of following advice.

While absolute helpfulness ratings showed only small mean differences, comparative evaluations revealed consistent and statistically significant advantages for conversation-based personalization. These conclusions remained significant after accounting for repeated evaluations within participants and applying FDR correction. In the experimental group (baseline = background-based personalization), conversation-based personalization was rated as significantly more helpful (FDR p-value $= 2.62\mathrm{E}{-04}$) and significantly more motivating in terms of likelihood of following advice (FDR p-value $= 4.20\mathrm{E}{-03}$). In the control group (baseline = no personalization), results similarly favored conversation-based personalization, with significant improvements in both perceived helpfulness (FDR p-value $= 3.45\mathrm{E}{-02}$) and likelihood of following advice (FDR p-value $= 1.75\mathrm{E}{-02}$). These results indicate that incorporating interaction history consistently improves participants' comparative evaluations of cybersecurity advice, even after accounting for repeated measurements within participants.

We also analyzed answer length across the subset of questions evaluated in the final survey. Personalized responses were slightly longer than non-personalized responses, though effect sizes were small ($d \leq 0.37$). Importantly, there was no meaningful difference in length between background-based and conversation-based personalization ($p = 0.21$), with a negligible effect size ($d \approx 0$). These findings suggest that the improvements observed from conversation-based personalization are not driven by increased verbosity, but rather reflect qualitative differences in how responses are tailored to users.

\begin{table*}[h]
\centering
\small
\renewcommand{\arraystretch}{1.2}
\begin{tabular}{|p{0.35\linewidth}|p{0.07\linewidth}|p{0.07\linewidth}|p{0.12\linewidth}|p{0.1\linewidth}|p{0.09\linewidth}|}
\hline
\textbf{Metric (Final Survey)} & \textbf{Base Mean} & \textbf{Pers. Mean} & \textbf{Evaluations} & \textbf{Cohen's d} & \textbf{FDR p-value} \\
\hline

\multicolumn{6}{|l|}{\textbf{Experimental Group (Base = Background Personalization, P1)}} \\
\hline
Helpfulness (1--4 scale) & 3.59 & 3.65 & 175 & 0.099 & 7.81e-01 \\
\hline
\cellcolor{lightgreen} Rel. Helpfulness (-2 - 2) 
& \multicolumn{2}{c|}{\cellcolor{lightgreen} 3.49e-01} 
& \cellcolor{lightgreen} 175 
& \cellcolor{lightgreen} 0.697 
& \cellcolor{lightgreen} 2.62e-04 \\
\hline
\cellcolor{lightgreen} Rel. Likelihood of Following (-2 - 2) 
& \multicolumn{2}{c|}{\cellcolor{lightgreen} 2.60e-01} 
& \cellcolor{lightgreen} 169 
& \cellcolor{lightgreen} 0.557 
& \cellcolor{lightgreen} 4.20e-03 \\
\hline

\multicolumn{6}{|l|}{\textbf{Control Group (Base = No Personalization)}} \\
\hline
Helpfulness (1--4 scale) & 3.77 & 3.75 & 110 & -0.131 & 5.62e-01 \\
\hline
\cellcolor{lightgreen} Rel. Helpfulness (-2 - 2) 
& \multicolumn{2}{c|}{\cellcolor{lightgreen} 1.45e-01} 
& \cellcolor{lightgreen} 110 
& \cellcolor{lightgreen} 0.538 
& \cellcolor{lightgreen} 3.45e-02 \\
\hline
\cellcolor{lightgreen} Rel. Likelihood of Following (-2 - 2) 
& \multicolumn{2}{c|}{\cellcolor{lightgreen} 1.64e-01} 
& \cellcolor{lightgreen} 110 
& \cellcolor{lightgreen} 0.570 
& \cellcolor{lightgreen} 1.75e-02 \\
\hline

\end{tabular}
\caption{Final survey evaluation of conversation-based personalization (P4). Base refers to the answer originally received during the study: background-based personalization (P1) in the experimental group and no personalization in the control group. Personalized refers to the alternate conversation-based personalized answer shown in the final survey. Helpfulness ratings were measured on a 1-4 scale. Comparative ratings were measured on a $-2$ to $2$ scale, where positive values favor the personalized answer. Statistical significance was assessed using analyses that account for repeated measurements within participants (linear mixed-effects models with participant-level random intercepts, with participant-level aggregation used when the mixed-effects model produced a singular fit). Reported p-values were adjusted using the Benjamini--Hochberg procedure across all final-survey hypothesis tests. The experimental and control groups contained 35 and 22 participants, respectively; ``Evaluations'' reports the number of question-level ratings contributing to each outcome. Cohen's $d$ values were computed from participant-level mean outcomes after averaging repeated evaluations within each participant.}
\label{tab:final_survey_results_p4}
\end{table*}

\paragraph{Consistency Across LLM-Based and Human Evaluations.}
To examine the correspondence between the two evaluation methods, we compared the two personalization contrasts represented in both settings: P4 versus the PMT-only baseline and P4 versus P1. Both contrasts favored P4 in the human evaluation for relative helpfulness and likelihood of following advice, and in the LLM-based evaluation across understandability, actionability, and motivating power. The relative ordering was also identical across methods: the improvement of P4 over P1 was larger than its improvement over the PMT-only baseline for every outcome. Thus, the two evaluations exhibited complete directional consistency across the shared contrasts (Appendix Table~\ref{tab:cross_method_directional_consistency}).

\subsection{Personalization Effects: A Few Examples}
\label{sec:qual_examples}

To better understand how conversation-based personalization improves response effectiveness, we examined representative examples of questions asked by users and answers generated under different conditions. Across these examples, conversation-based personalization improved answer quality by adapting to user constraints, leveraging prior experiences, and providing more targeted guidance. The example questions along with full answers generated with no personalization, background-based personalization, and conversation-based personalization are provided in Appendix \ref{sec:ex_ans_appendix}.

Several recurring patterns emerged. First, conversation-based personalization adapted recommendations to users' existing tools and behaviors. For example, responses incorporated information about previously discussed security tools and workflows, making recommendations easier for users to understand and implement. Second, recommendations were grounded in prior experiences, increasing the relevance of the advice. Finally, interaction history enabled the system to infer user intent and provide more targeted guidance, reducing overhead required for the user and enabling more immediate action.

The improvements observed tended to depend on behavioral signals derived from interaction history rather than self-reported attributes. These qualitative patterns help explain the quantitative results presented earlier, where conversation-based personalization consistently outperformed both non-personalized and background-based approaches.

\subsection{Relevance of User Questions to Real-World Context}
\label{sec:authenticity_user_questions}

We evaluated the extent to which participants' questions reflected their own real-world concerns, rather than being asked primarily to satisfy the study requirement of submitting at least two questions per day. Overall, participants reported that most questions were related to real security issues they encountered, while relatively few indicated that their questions were primarily driven by the study requirement (see Appendix \ref{sec:rel_to_real_world_appendix} for detailed results).

\begin{figure}[H]
    \centering
    \includegraphics[width=0.8\textwidth]{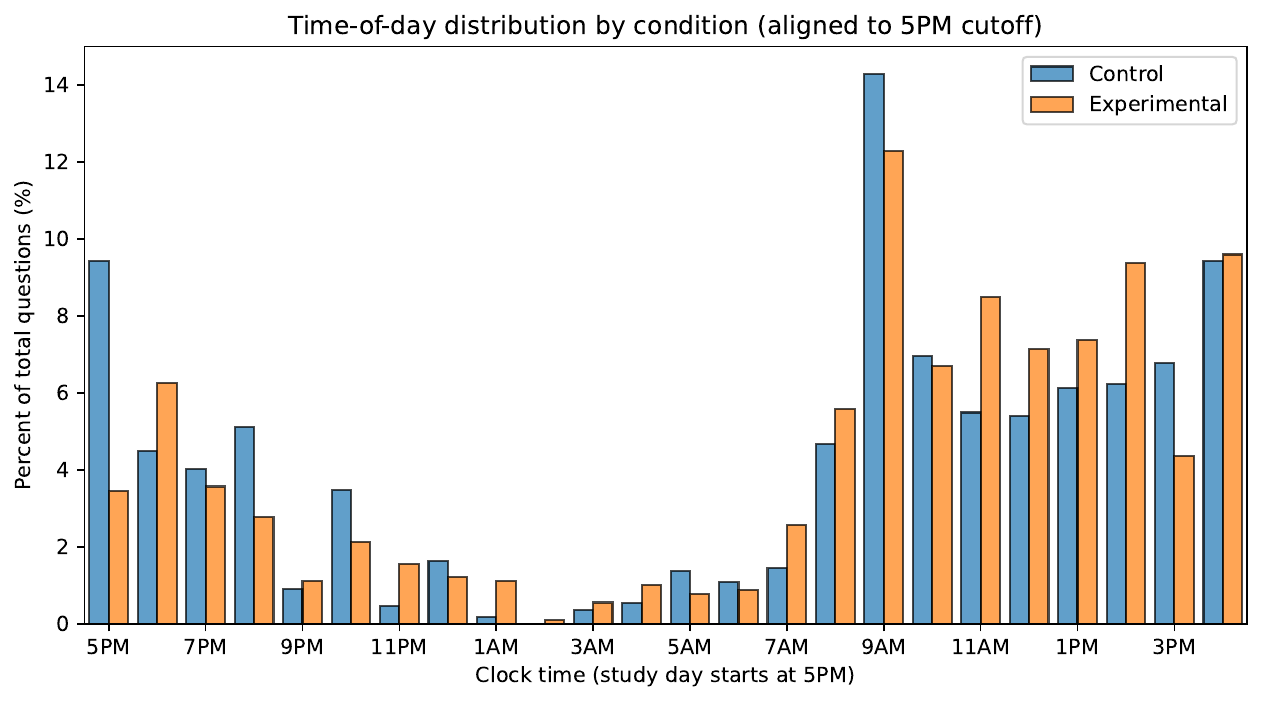}
    \caption{Time-of-day distribution of participant questions by condition (control vs. experimental), aligned to the 5 pm Eastern Standard Time daily cutoff (questions after 5 pm are counted toward the next study day). Bars show the percent of total questions within each condition. Besides sleeping hours, the times at which participants asked questions were spread throughout the day.}
    \label{fig:timeofday}
\end{figure}

Behavioral patterns further support this interpretation. As shown in Figure~\ref{fig:timeofday} (Appendix \ref{sec:question_time_dist_appendix}), questions were distributed throughout the day rather than clustered immediately before the daily cutoff, suggesting that participants engaged with the assistant as questions arose naturally.

We also categorized questions using the taxonomy developed by \citet{duesterwald2025}. Questions spanned a broad range of cybersecurity domains, indicating that participants engaged with diverse and practical security concerns (see Appendix \ref{sec:q_topic_dist_appendix}).

\subsection{User Experience and Behavioral Insights}
\label{sec:qualitative_insights}

We complement these quantitative findings with self-reported outcomes and qualitative feedback from the final survey and follow-up interviews. Responses regarding continued access (Appendix \ref{sec:sustained_intrest_appendix}) indicate strong sustained interest in the tool, with only a small fraction of participants expressing no interest in continued use.

Participants reported frequently using the assistant to address security questions encountered in their daily lives, including issues they had previously postponed addressing. Participants also emphasized that a primary value of the assistant was its ability to provide clear explanations and communicate why recommended actions were important.

Although users reported high levels of understanding and perceived usefulness, translating advice into action was strongly influenced by the effort required. Lower-effort behaviors were frequently adopted immediately, whereas more time-intensive recommendations were often deferred for later implementation.

\begin{quote}
"Anything I could do quickly (cache clearing) I did. But I also copy and pasted several paragraphs of advice to follow later when I'm not so busy...More time consuming procedures will have to wait a bit, and even the tool said you can't do it all overnight. I thought that was affirming."

\end{quote}
Overall, these findings suggest that participants used the assistant to address real-world concerns and integrated its recommendations into their behavior. More generally, participants reported greater awareness of the importance of security practices and viewed more involved recommendations as worthwhile, even when implementation was postponed.

\section{Discussion}

Across both our automated LLM-based evaluations and in situ user study, we observe a consistent pattern: conversation-based personalization, which adapts responses using prior interaction history, yields the most reliable improvements in user-perceived effectiveness. In contrast, static background-based personalization produces mixed or limited gains relative to a strong PMT-based baseline. Taken together, these results suggest that how personalization is implemented matters substantially, and that behavior-driven personalization is more effective than static user profiling in this domain.

\subsection{From LLM-Based Evaluation to Behavioral Personalization}

A key contribution of this work is demonstrating alignment between LLM-based evaluation and real-world user outcomes. In the LLM evaluations, conversation-based personalization consistently outperformed other approaches across understandability, actionability, and motivating power. The in situ study reproduced the same directional pattern: participants' comparative ratings significantly favored conversation-based answers in both helpfulness and likelihood of following the advice. Although absolute differences in helpfulness ratings were modest, comparative judgments consistently favored conversation-based personalization and remained statistically significant after accounting for repeated evaluations within participants.

This convergence across methodologies strengthens confidence that the observed improvements are not artifacts of model-based evaluation, but instead reflect meaningful differences in how users perceive and respond to advice. More broadly, these findings suggest that LLM-based evaluation can provide a scalable mechanism for comparing personalization strategies before undertaking costly user studies.

Our findings also highlight an important distinction between two forms of personalization. Background-based personalization, which relies on static survey responses, showed limited benefits. In contrast, conversation-based personalization leverages behavioral signals derived from interaction history, enabling responses that adapt to users' evolving concerns and preferences. By conditioning on observed behavior rather than self-reported characteristics, these systems can capture aspects of user context that are difficult to elicit through static surveys alone. The qualitative examples presented in Section~\ref{sec:qual_examples} illustrate how interaction history enables systems to adapt recommendations to users' existing tools and behaviors, ground advice in prior experiences, and provide more targeted guidance.

Together, these results suggest that personalization in cybersecurity assistants should move beyond static user profiles toward adaptive, interaction-driven approaches. Because deployed assistants naturally accumulate interaction logs, follow-up questions, and feedback, this form of personalization can be supported in real-world settings without requiring additional instrumentation.

\subsection{Privacy and Human-Centered Design Considerations}

While richer behavioral signals may enable more effective personalization, they also raise important questions surrounding privacy, consent, and data use. The interaction histories used in our study were voluntarily generated through users' conversations with the assistant, but future systems may have access to much richer behavioral context.

Realizing the benefits of behavior-driven personalization will require mechanisms that provide users with meaningful transparency and control over how their data are used. In domains such as cybersecurity, where the goal is to influence behavior, balancing personalization with user agency and privacy may be central to designing trustworthy assistants.

\subsection{Limitations}

Our evaluation focuses on user perceptions and self-reported likelihood of following security advice rather than direct observation of real-world behavior. Although the LLM-based evaluation provides a useful mechanism for comparing personalization strategies at scale, it relies on an LLM-based judge. We partially mitigate this limitation through a real-world user study, but additional forms of human evaluation could further strengthen this connection.

The study duration was limited to seven days, constraining the amount of interaction history available for personalization. Differential dropout produced an imbalance between the experimental ($n=35$) and control ($n=22$) groups, reducing statistical power for the nonsignificant between-group comparison of absolute-helpfulness improvement (Table~\ref{tab:final_survey_results_p4}). The primary final-survey analyses account for repeated evaluations within participants using mixed-effects models, but the modest participant sample remains a limitation. Finally, we do not measure long-term behavioral change, and understanding how personalization influences sustained security practices remains an important direction for future work.

\section{Conclusion}

In this work, we investigated personalization as a mechanism for enhancing the effectiveness of LLM-powered cybersecurity assistants. We first conducted a large-scale automated LLM-based evaluation over a corpus of real-world cybersecurity questions, which identified personalization based on accumulated interaction history as a particularly promising approach. We then validated these findings through a 7-day in situ user study involving 57 participants and 1,066 questions. Across both studies, conversation-based personalization consistently improved perceived helpfulness and self-reported likelihood of following security advice in comparative evaluations, outperforming static background-based approaches and yielding gains beyond an already strong PMT-based baseline.

Taken together, these findings suggest that behavior-driven personalization provides a more effective foundation for cybersecurity assistants than static user profiling. More broadly, the agreement between LLM-based evaluation and human evaluation suggests that LLM-based evaluations may provide a scalable mechanism for comparing personalization strategies before undertaking costly user studies.

As language-model systems become increasingly personalized, understanding how to leverage behavioral context while preserving user privacy and agency will become increasingly important. Our results suggest that conditioning on observed user behavior offers a promising mechanism for narrowing the gap between providing security advice and encouraging users to act on it.

\section{Acknowledgments}
This research has been supported in part by grants from the National Science Foundation, including grants under the SaTC program (grant CNS-1914486), under the Smart and Connected Communities (grant 2426911) and under the REU program (grant 2150217). Additional funding was provided by Meta and by the Block Center for Technology and Society.

\bibliographystyle{colm2026_conference}
\bibliography{colm2026_conference}

\clearpage
\appendix
\FloatBarrier
\section{Appendix}

\subsection{Prompts Used for Automated Personalization Evaluation}
\label{sec:eval_prompts_appendix}

\begin{table}[H]
\centering
\small
\renewcommand{\arraystretch}{1.2} 
\begin{tabular}{|p{0.11\linewidth}@{\hspace{0.2cm}}|p{0.83\linewidth}@{\hspace{0.2cm}}|}
\hline

\multicolumn{2}{|l|}{\textbf{Prompts Used to Generate Personalized Answers}} \\
\hline

{ \textbf{Type} }& \textbf{Prompt}\\
\hline

{ \textbf{P1} (PMT + static background-based personalization with 6 questions)}& I have a question I am hoping you can help me with: + \textit{question} + . In answering my question, \sethlcolor{yellow!30} \hl{please consider my answers to the following 6 questions and personalize your own answer to make it as effective as possible} based on what you can infer about me from my answers + \textit{context} + . In answering this question, please keep in mind that I am not a technical expert. \sethlcolor{blue!20}\hl{If your answer includes recommendations or warnings, please make sure to help me understand the risks of not heeding your advice and how critical this is}. Please limit your answer to no more than 250 words.\\
\hline

{ \textbf{P2} (PMT + static background-based personalization with 56 questions)}&I have a question I am hoping you can help me with: + \textit{question} + . In answering my question, \sethlcolor{yellow!30}\hl{please consider my answers to the following 56 questions and personalize your own answer to make it as effective as possible} based on what you can infer about me from my answers + \textit{context} + . In answering this question, please keep in mind that I am not a technical expert. \sethlcolor{blue!20}\hl{If your answer includes recommendations or warnings, please make sure to help me understand the risks of not heeding your advice and how critical this is}. Please limit your answer to no more than 250 words.\\
\hline

{ \textbf{P3} (PMT + personalization with few-shot examples based on tech level)}& In answering this question:  + \textit{question}  +  \sethlcolor{yellow!30}Please keep in mind that I have rated my comfort with cybersecurity technologies as around + \textit{comfort$\_$rating} +  on a scale of 1-5 so \hl{please customize your answer to be most appropriate to my tech background so I can understand it. Here are some examples of the type of customization that would help me best understand your answer + \textit{few$\_$shot}} + In answering this question, please keep in mind that I am not a technical expert. \sethlcolor{blue!20}I\hl{f your answer includes recommendations or warnings, please make sure to help me understand the risks of not heeding your advice and how critical this is.}
\\
\hline

{ \textbf{P4} (PMT + conversation-history based personalization)} & In answering this question: + \textit{question} + Please keep in mind this context of my previous questions, answers, alternate answers, and how I ranked them. + \textit{context} +  \sethlcolor{yellow!30}\hl{Please use that context to learn the kind of response I prefer and tune your response to be the best for me.} In answering this question, please keep in mind that I am not a technical expert. \sethlcolor{blue!20}\hl{If your answer includes recommendations or warnings, please make sure to help me understand the risks of not heeding your advice and how critical this is.}
\\
\hline

\end{tabular}
\caption{The prompts used to generate personalized answers for the four personalization strategies. Type refers to the personalization type. In the prompt, '+' indicates text that was appended, this includes the user question (\textit{question})}
\label{tab:personalization_ans_gen_prompts}
\end{table}

\clearpage
\begin{table}[H]
\centering
\small
\renewcommand{\arraystretch}{1.2} 
\begin{tabular}{|p{0.06\linewidth}@{\hspace{0.2cm}}|p{0.87\linewidth}@{\hspace{0.2cm}}|}
\hline

\multicolumn{2}{|l|}{\textbf{Prompts Used to Generate Personalized Answers}} \\
\hline

{ \textbf{Type} }& \textbf{Prompt}\\
\hline

{ \textbf{Multi} }&        Compare the following two answers to the following question: + \textit{question} + 
        In your comparison, consider that a good answer is one that is accurate, easy to understand, actionable (if a recommendation is provided) and that motivates the recipient to follow any recommendation included in the answer.
        Here are the two answers to compare (A and B):
        Answer A: + \textit{answer} + 
        Answer B: + \textit{answer2} + 
        Your comparison should be in the form of an integer between +2 and -2 and should successively provide an integer for answer quality based on (a) accuracy, (b) understandability, (c) actionability, (d) motivating power.
        2 indicates that Answer B is significantly better than Answer A;
        1  indicates that Answer B is somewhat better than Answer A;
        0: The overall quality of the two answers is about the same. Neither answer seems to be better than the other;
        -1  indicates that Answer A  is somewhat better than Answer B;
        -2  indicates that Answer A is significantly better than Answer B;
        Your answer should be in the following format:
        (a) Accuracy: enter an integer between -2 and +2
        (b) Understandability: enter an integer between -2 and +2
        (c) Actionability: enter an integer between -2 and +2
        (d) Motivating power: enter an integer between -2 and +2
        (e) Overall score: enter an integer between -2 and +2

\\
\hline

{ \textbf{Ind} } & Evaluate the following answer to a cybersecurity question. In your evaluation consider that a good answer is one that is accurate, easy to understand, actionable (if a recommendation is provided) and that motivates the recipient to follow any recommendation included in the answer. Here is the question: + \textit{question} + and here is the answer to evaluate: + \textit{answer} +. Your evaluation should provide a score on a scale of 1 to 10 - where a score of 1 indicates an answer that might be inaccurate and does not motivate a user to follow their advice at all, a score of 5 indicates an answer that might provide an accurate answer but might be phrased slightly confusingly and take no steps to motivate the user who asked the question, and a score of 10 indicates an answer that is accurate, provides clear next steps that are easy to follow, and highlights risks to really motivate the user to follow the advice and take steps to protect themselves. "Please provide an score from 1-10 along 5 metrics, (a) accuracy (here rank the answer from 1-10 based on how accurate the information is provided), (b) understandability (here rank the answer from 1-10 based on whether the average internet user would be able to understand this answer), (c) actionability (if the answer provides advice, please rank the answer from 1-10 based on how easy and clear it is to follow the advice provided), (d) motivating power: (here rank the answer from 1-10 based on whether the answer really motivates a user to follow the advice and take action, perhaps by highlighting the risks), and (e) an overall score (here rank the answer from 1-10 on overall effectiveness which combines all metrics. Make sure to format your answer in the following way: (a) Accuracy: [accuracy score 1-10] (b) Understandability: [understandability score 1-10] (c) Actionability: [understandability score 1-10] (d) Motivating power: [motivating power score 1-10] (e) Overall score: [overall score 1-10]"
\\
\hline

\end{tabular}
\caption{The prompts used to evaluate personalized answers. Type refers to the type of comparison that is being evaluated. "multi" indicates two different answers to the same question are being compared, "ind" indicates an individual answer is being rated. In the prompt, '+' indicate text that was appended. This includes the user question (\textit{question}), and one or more answer, \textit{answer}, and \textit{answer2}}
\label{tab:personalization_ans_eval_prompts}
\end{table}

\FloatBarrier
\clearpage

\clearpage
\subsection{Full Context for Personalization Evaluation}
\label{sec:full_context_pers_appendix}
\vspace{12pt}

\definecolor{lightgray}{RGB}{240,240,240}
\setlength{\dashlinedash}{0.5pt}   
\setlength{\dashlinegap}{2pt}      
\setlength{\arrayrulewidth}{0.4pt} 

\newcolumntype{P}{>{\raggedright\arraybackslash}p{0.96\textwidth}}

\begin{longtable}{|P|}
\hline
\textbf\underline{{P3, Few-Shot Examples}} \\
\hline

\textbf{Low Comfort Level} \\
\hline
\textit{I have a relatively low level of comfort with cybersecurity tech so I would benefit from more explanation and avoiding too many technical terms/jargon that might make me unable to answer the question. Here are some examples of questions and the types of answers I would want:} \\
Example question 1: What is ransomware, and how can I protect myself from it?, Example answer 1: Ransomware is a kind of malicious software (a “virus”) that locks your computer or files so you can’t use them until you pay money (a “ransom”) to the attacker. Usually, it spreads through links or attachments in fake emails or unsafe websites. To protect yourself: - Don’t click on unknown links or attachments in emails or texts. - Keep your computer updated so it can block new threats. - Back up your files on an external drive or cloud service, so you can restore them if attacked. - Install antivirus software and let it run automatic scans. If you ever see a message demanding payment to unlock your files, don’t pay it—contact a tech expert or your IT department right away. Please feel free to ask for any clarification or other question you may have. \\
Example question 2: What are the signs of a compromised device, and what should I do if I suspect a breach?, Example answer 2: A compromised device means someone or something harmful might have gained access to it. You might notice signs like: - The device suddenly feels much slower than usual. - You see pop-ups or ads that weren’t there before. - Apps open or close on their own. - The battery drains quickly or your data usage is higher than normal. - Your friends say they’re getting strange messages or emails from you. If you think your device might be compromised: - Disconnect from the internet (turn off Wi-Fi and mobile data). - Run a full antivirus or security scan using your phone or computer’s built-in security app or a trusted one (like Norton, McAfee, or Windows Defender).- Change your passwords — start with email and banking accounts. - Update your software — hackers often exploit old versions. - Contact support if you’re not sure what to do — for example, your phone company or local computer repair service. Would you like me to explain how to do any of these steps in more detail?\\
Example question 3: I am joining a public wifi network I don't trust. What are some steps I can take to stay safe?, Example answer 3: If you’re connecting to a public Wi-Fi network that you don’t fully trust — for example, at a coffee shop or airport — there are a few simple things you can do to protect yourself: Avoid entering personal information (like passwords or credit card numbers) while on that network if you can. Turn off file sharing and AirDrop so others can’t access your files. Use a VPN (Virtual Private Network) if possible — it scrambles your internet traffic so others on the same Wi-Fi can’t easily see what you’re doing. Only visit secure websites — look for https:// at the start of the web address (that “s” means it’s using encryption). Turn off the Wi-Fi when you’re done so your device doesn’t reconnect automatically later. Would you like me to explain how to use a VPN or check if your device has sharing turned off?\\

\hline
\textbf{Medium Comfort Level}  \\
\hline 
\textit{I have a medium level of comfort with cybersecurity tech so keep your answers relatively detailed with some tech terms but nothing too complicated. Here are some examples of questions and the types of answers I would want:}\\
Example question 1: What is ransomware, and how can I protect myself from it?, Example answer 1: Ransomware is a type of malware that encrypts your files or locks you out of your computer until you pay a ransom to the attacker. It often spreads through phishing emails, malicious downloads, or compromised websites. You can protect yourself by: - Being cautious with email attachments and links—only open them if you’re sure they’re safe. - Keeping your operating system and software up to date—updates often patch security flaws that ransomware exploits. - Using reliable antivirus and anti-malware software to detect and block threats. - Backing up your data regularly to an external drive or cloud storage that’s not always connected to your computer.  If you do get infected, disconnect from the internet and contact IT support or a cybersecurity professional—don’t pay the ransom unless directed by experts. Please feel free to ask any follow up questions you may have.\\
Example question 2: What are the signs of a compromised device, and what should I do if I suspect a breach?, Example answer 2: A device may be compromised if it behaves abnormally — for example: - You notice new apps, extensions, or files you didn’t install. - Your browser homepage or search engine changes without permission. - CPU or network usage is unusually high even when idle. - You’re redirected to strange websites or your accounts show logins from unknown locations. If you suspect a breach: - Disconnect from the internet to prevent further data exfiltration. - Run a reputable antivirus or malware scan (e.g., Windows Defender, Malwarebytes, or Bitdefender). - Change all important passwords, preferably from another clean device. - Check account activity (email, bank, social media) for unauthorized logins. - Apply OS and app updates and remove suspicious software. If it still behaves oddly, back up your files and consider a factory reset. Would you like me to go over how to run a malware scan or what tools  
to use?\\
Example question 3: I am joining a public wifi network I don't trust. What are some steps I can take to stay safe?, Example answer 3: When connecting to an untrusted  public Wi-Fi, your main goal is to reduce the chance of others intercepting or tampering with your data. Here are a few good practices: Use a reputable VPN service — this encrypts your network traffic end-to-end. - Stick to HTTPS websites — a browser extension like HTTPS Everywhere can enforce that. - Turn off file and printer sharing, and disable automatic connections to open networks. - Avoid logging into sensitive accounts (like banking) unless you’re using your mobile data instead. - Keep your system and apps updated — updates often fix known security holes. - Use two-factor authentication (2FA) so even if your password leaks, attackers can’t log in easily. Do you want me to  
recommend specific VPNs or settings for your operating system? \\

\hline
\textbf{High Comfort Level} \\
\hline

\textit{I have a relatively high level of comfort with cybersecurity tech so please include a lot of details so I get as much information out of your answer as possible. Here are some examples of questions and the types of answers I would want:}\\
Example question 1: What is ransomware, and how can I protect myself from it?, Example answer 1: Ransomware is a form of malicious code that encrypts user data using symmetric or asymmetric cryptography (often AES + RSA). Attackers demand payment—typically in cryptocurrency—in exchange for the decryption key. Common variants include LockBit, BlackCat (ALPHV), and Clop, which propagate through phishing campaigns, Remote Desktop Protocol (RDP) brute-forcing, or software supply-chain attacks. Protection strategies include: Implementing the 3-2-1 backup rule (three copies of data, on two different media, one off-site/offline). Maintaining up-to-date endpoint protection and EDR/XDR tools with real-time behavioral monitoring. Disabling macros and PowerShell where not needed, since many ransomware droppers use these for execution. Applying least-privilege principles and network segmentation to limit lateral movement. Regularly patching OS and third-party software to eliminate vulnerabilities exploited by ransomware loaders (e.g., via CVE-chaining). If compromised, isolate infected hosts, preserve forensic evidence, and engage incident response teams—never assume decryption tools from attackers are trustworthy. Please feel free to ask follow up questions if you would like more details or have any other questions.\\
Example question 2: What are the signs of a compromised device, and what should I do if I suspect a breach?, Example answer 2: Indicators of compromise (IoCs) can include both behavioral and forensic signs: - Unexpected outbound network traffic, unusual DNS queries, or persistent C2 connections. - Elevated CPU/GPU usage without active processes (possible cryptojacking). - Unauthorized privilege escalations or unknown scheduled tasks/services. - Altered system files, registry keys, or rootkit-like persistence mechanisms. - Browser injection, credential theft attempts, or suspicious PowerShell execution logs. If a breach is suspected: - Isolate the system from the network immediately. - Collect volatile data (network connections, running processes, RAM dump) before rebooting. - Run endpoint detection or forensic tools like Sysinternals, Volatility, or CrowdStrike Falcon to identify anomalies. - Preserve logs for incident response and threat attribution. - Change credentials and reissue API keys/tokens. - Reimage or restore from a trusted backup after root cause analysis. Would you like me to go deeper into recommended forensic tools or incident-response workflows?\\
Example question 3: I am joining a public wifi network I don't trust. What are some steps I can take to stay safe?, Example answer 3: When using an untrusted public Wi-Fi, assume the local network could be hostile — e.g., susceptible to MITM (man-in-the-middle) attacks, rogue DHCP servers, or traffic sniffing. Here’s a more robust approach: - Always use a trusted VPN with strong encryption (AES-256 or ChaCha20) and DNS leak protection. - Enforce HTTPS/TLS — use HSTS-preloaded browsers or extensions to prevent downgrade attacks. - Disable all unnecessary network services (SMB, AirDrop, discovery protocols). - Manually verify SSID authenticity — attackers can spoof hotspot names to run evil twin attacks. - Use personal hotspots or tethering when feasible for sensitive logins. - Ensure your OS firewall is active and block inbound connections. - Use MFA/2FA and consider hardware tokens (like YubiKey) for critical accounts. - Monitor for ARP spoofing or DNS poisoning if you have packet-capture tools like Wireshark or Little Snitch. Would you like me to go deeper into secure VPN configuration or Wi-Fi attack detection tools?\\

\hline
\hline
\rowcolor{lightgray}
\textbf\underline{{P4, Accumulated Conversation Context}} \\
\hline
\rowcolor{lightgray}
\noindent MY QUESTION:  + \textit{user question} + \\
\rowcolor{lightgray}
THE ANSWER I RECEIVED: + \textit{base chatbot answer} + \\
\rowcolor{lightgray}
AN ALTERNATE ANSWER I COULD HAVE RECEIVED: + \textit{alternate answer}\\
\rowcolor{lightgray}
MY RATINGS FOR THIS ANSWER + ALTERNATIVE ANSWER:\\
\rowcolor{lightgray}
Q: How many terms were there in the answer that were confusing or difficult to understand (1 = no terms, 3 = a couple, 5 = enough confusing terms that the answer was not understandable) $--> A$: + \textit{user confusing terms rating} + \\
\rowcolor{lightgray}
Q: Overall, how well were you able to understand the answer? (1 = not at all, 5 = I completely understood the answer) $-->$ A: + \textit{user understandability rating} + \\
\rowcolor{lightgray}
Q: How easy was it to follow the advice given (if instructions were given)? (1 = impossible to follow, 5 = very easy to follow) $-->$ A: + \textit{user follow advice rating} + \\
\rowcolor{lightgray}
Q: Did you follow the advice provided by the assistant? $-->$ A:  + \textit{user followed advice answer} + \\
\rowcolor{lightgray}
Q: Why or why not? Please explain. $-->$ A:  + \textit{user text answer} + \\
\rowcolor{lightgray}
Q: Do you plan to follow the advice in the future? $-->$ A: " + \textit{user plan follow rating} + \\
\rowcolor{lightgray}
Q: On a scale of 1 - 4 how would you rate the answer you received to this answer? (1 = not helpful at all, 2 = not terribly helpful 3 = somewhat helpful, 4 = very helpful) $-->$ A:  + \textit{user alternate answer rating} + \\
\rowcolor{lightgray}
Q: If you had received the above alternate answer would you have found it more or less helpful? $-->$ A: + \textit{user comparing alternate answer helpfulness rating} + \\
\rowcolor{lightgray}
Q: If you had received the above alternate answer would you have been more or less likely to follow the advice provided (if applicable)? $-->$ A: + \textit{user more likely to follow alternate answer rating} + \\
\hline
\caption{The full contents of prompts used for personalization. For P3 (few-shot prompting with three technology knowledge levels), the full few-shot examples for each technology level are provided. For P4 (accumulated conversation context and user evaluation), the full accumulation of user questions, chatbot answers, and user evaluations is provided.}
\label{tab:full_context_pers}
\end{longtable}

\clearpage
\subsection{Example Answers}
\label{sec:ex_ans_appendix}
\vspace{12pt}

\definecolor{lightgray}{RGB}{240,240,240}
\setlength{\dashlinedash}{0.5pt}   
\setlength{\dashlinegap}{2pt}      
\setlength{\arrayrulewidth}{0.4pt} 

\begin{longtable}{|P|}
\hline
\textbf{Example Question 1: } What are the recommended requirements for a strong password? \\
\hdashline

\textbf{Answer with no personalization:} \\

A strong password is essential for protecting your online accounts from hackers. Here’s what experts recommend for a strong password, explained in simple terms:

**1. Make it long:**  
Your password should be at least 12 characters. Longer passwords are much harder for hackers to guess.

**2. Mix it up:**  
Use a combination of uppercase letters (A, B, C), lowercase letters (a, b, c), numbers (1, 2, 3), and symbols ($!, @, \#, \$$). This variety makes your password less predictable.

**3. Avoid personal info:**  
Don’t use names, birthdays, or common words (like “password” or “123456”). Hackers often try these first.

**4. Use unique passwords:**  
Never use the same password for multiple accounts. If one account is hacked, others could be at risk if they share the same password.

**Why it matters:**  
If you use weak or repeated passwords, hackers can easily break into your accounts using automated tools. This could lead to stolen money, identity theft, or loss of important personal data. Many people don’t realize they’ve been hacked until after damage is done.

**What you can do:**  
Consider using a password manager (a secure app that creates and remembers strong passwords for you). This makes it much easier to have strong, unique passwords for every account.

**Protecting your accounts with strong passwords is a simple but critical step to keep your online life safe.**

\\
\hdashline

\textbf{Answer with background-based personalization:} \\
A strong password is a vital line of defense for your privacy and security online. Even if you’re careful about links and downloads, a weak password can make it easy for hackers to access your accounts—often without you realizing it. Here’s what’s recommended for a strong password:

- **Length:** Use at least 12 characters. Longer is better.
- **Complexity:** Combine uppercase and lowercase letters, numbers, and special symbols (like $!, @, \#, \$$, etc.).
- **Unpredictability:** Avoid common words, names, or predictable patterns like “password123” or “qwerty.”
- **Uniqueness:** Never reuse passwords across different accounts. Each account should have its own, unique password.

**Why this matters:**  
Even if you’re vigilant about privacy and use tools to protect your browsing, a weak or reused password is an easy way for attackers to break in. Criminals use automated tools to guess passwords quickly, and once they’re in, they can steal your data, impersonate you, and potentially bypass other defenses you’ve set up. Many breaches happen silently—you may not notice until significant damage is done.

**What you can do:**  
Consider using a reputable password manager. It will generate and remember strong, unique passwords for each account, so you don’t have to. This is a practical step that adds a strong layer of security and supports your goal of maintaining your privacy online.
\\
\hdashline

\textbf{Answer with conversation-based personalization:} \\
**Recommended requirements for a strong password (simple + practical):**

1. **Make it long (most important):**  
   Aim for **14–20+ characters**. Longer passwords are much harder for attackers to crack, even with powerful computers.\\
2. **Use a random password (best option):**  
   Since you use **Bitwarden**, the easiest and strongest approach is: **let it generate a random password** (a mix of letters/numbers/symbols). This avoids “guessable” patterns humans tend to use.\\
3. **Or use a passphrase you can remember (good option):**  
   Use **4–6 random words** strung together (and you can add a number/symbol). Example style: `river-lamp-planet-salsa-9`.  
   This is usually easier to type and still very strong if the words are truly random.\\
4. **Unique for every account:**  
   Never reuse passwords. If one site is breached, attackers try the same password on your email, banking, etc. This is one of the most common ways people get hacked.\\
5. **Avoid predictable content:**  
   Don’t use names, birthdays, pet names, keyboard patterns (like `qwerty`), or common substitutions (`P@ssw0rd`). Attackers try these first.

**How critical is this?**  
Very. Weak or reused passwords can lead to **account takeovers**, **financial theft**, and **identity fraud**, often with little warning.

**Extra strong step:** Turn on **2-factor authentication (2FA)** for important accounts (especially email and your password manager).
\\
\hline
\hline
\rowcolor{lightgray}
\textbf{Example Question 2: } Which is more secure to my online banking with, my iMac or my smartphone? \\
\hdashline
\rowcolor{lightgray}
\textbf{Answer with no personalization:} \\
\rowcolor{lightgray}
Great question! Both your iMac and smartphone can be secure for online banking, but your *smartphone* is generally safer—**if** you follow some basic precautions.

**Why a smartphone is usually safer:**
- **Isolation:** Smartphones are designed to keep apps separate, which helps prevent malware spreading.
- **Updates:** Phones often update automatically, fixing security issues quickly.
- **Fewer threats:** Most viruses and malware target computers, not phones.

**Best practices for smartphone banking:**
1. **Update your phone and apps regularly.** This patches security holes.
2. **Use your bank’s official app** (not the website via browser) for stronger security.
3. **Enable a screen lock and, if possible, fingerprint or face unlock.**
4. **Never use public Wi-Fi for banking.** Use your home Wi-Fi or cellular data.
5. **Don’t install apps from unknown sources.** Stick to the Apple App Store or Google Play.

**iMac risks:**
- Macs can get malware, especially if you download from untrusted sites or open suspicious links.
- Browsers can be targeted by phishing attempts (fake websites that steal your info).
- Security updates might be delayed if you ignore them.

**Risks of not following advice:**
- Your personal and banking data could be stolen.
- Thieves could move money out of your accounts.
- Fixing identity theft is stressful and time-consuming.

**Bottom line:** Your smartphone, used carefully, is usually the safer choice for online banking. Always keep it updated and use your bank’s official app. Don’t ignore these steps—they protect your money and identity!
\\
\hdashline
\rowcolor{lightgray}
\textbf{Answer with background-based personalization:} \\
\rowcolor{lightgray}
Based on your habits and comfort level with technology, I recommend using your **smartphone** for online banking, as it tends to offer a higher level of security for most people in your situation.

**Why your smartphone is safer:**  
Smartphones are designed with stronger security in mind, especially for everyday users. Your bank’s official app (which you should always use instead of the website) is isolated from other apps, making it much harder for a virus or hacker to access your banking information. Phones also update automatically more often, closing security gaps quickly.

**Risks of not following this advice:**  
If you use your iMac, it can be harder to spot if something is wrong—viruses and malware can hide in the background, and you might not notice if your information is being stolen. Hackers often target computers, and simply clicking on a deceptive link or downloading the wrong thing can put you at risk. If your banking details are compromised, someone could steal your money or your identity, which is very difficult and stressful to fix.

**How to stay safe on your smartphone:**  
- Keep your phone and apps up to date.  
- Use your bank’s official app, not a web browser.  
- Use a strong password or fingerprint/face lock.  
- Avoid doing banking on public Wi-Fi.

Taking these steps protects your money and your privacy. You don’t need to be technical—just being careful and using your phone’s built-in security features goes a long way.
\\
\hdashline
\rowcolor{lightgray}
\textbf{Answer with conversation-based personalization:} \\
\rowcolor{lightgray}
If both are fully updated and protected, **your smartphone is usually the more secure choice for online banking**—especially if you use your bank’s official app.\\
**Why the smartphone often wins**
- **Stronger “built-in” protection:** \\iPhones/Android phones are more locked down by design, and apps are more restricted from snooping on each other.  
- **Safer sign-in options:** Phones commonly use Face ID/Touch ID and can store **passkeys** more smoothly. This reduces the risk of password theft.
- **Less exposure to risky add-ons:** Computers are more likely to have browser extensions, plug-ins, or downloaded programs that can quietly interfere (you saw how an extension like Kudos can pop up during banking).

**When your iMac can be just as safe**
Your iMac can be very safe **if** you keep macOS, Safari/Chrome, and Malwarebytes updated, avoid unnecessary extensions, and only bank on trusted networks (like your home Wi‑Fi).

**Most important risks to avoid (on either device)**
- **Phishing (fake bank sites):** This is a top cause of account theft. Always type the bank URL yourself or use the app.
- **Public Wi‑Fi banking:** Higher risk of interception or trick networks. If you must, use cellular data or a trusted VPN.

**My practical recommendation**
For everyday banking: **use your smartphone + the official banking app + Face ID/Touch ID (or a passkey if offered).** Use the iMac for statements/printing, but keep the browser “clean” (minimal extensions).
\\
\hline
\hline
\textbf{Example Question 3: } Where can I change my privacy settings? \\
\hdashline

\textbf{Answer with no personalization:} \\
The place to change your privacy settings depends on the account or service you’re using (like Facebook, Google, or your email). Here’s how you can find privacy settings on most websites and apps:

**General Steps:**
1. **Log in** to your account.
2. **Find your profile icon** or your name—usually in the top right corner.
3. Click it, and look for options called **“Settings,” “Account,”** or **“Privacy.”**
4. Inside, look for a section labeled **“Privacy,” “Security,”** or **“Privacy $\&$ Security.”**  
5. From here, you can adjust who sees your information, what data is collected, and more.

**Why this matters:**  
Ignoring privacy settings can put your personal information at risk. For example, strangers could see your posts, your email could be shared, or your location could be tracked. This can lead to spam, scams, or even identity theft.

**Critical advice:**  
- **Set your profile to private** when possible.
- **Limit who can see your posts, contact info, and activities.**
- **Review connected apps** and remove ones you don’t use.
- **Update settings regularly** because platforms change their privacy features.

**If you need help:**  
Most services have easy guides or help centers—search for “[service name] privacy settings” (for example, “Facebook privacy settings”).

**Bottom line:**  
Updating your privacy settings is essential for protecting your personal information. It’s worth the few minutes it takes. 
\\
\hdashline

\textbf{Answer with background-based personalization:} \\
Absolutely, I’ll help you clearly and simply.

To change your privacy settings, start by logging into the website or app where you want more privacy (like Facebook, Google, or your email). Look for your profile icon or name, often in the top right corner. Click it, and choose “Settings” or “Account.” Inside, you’ll usually find a “Privacy” or “Privacy $\&$ Security” section—this is where you can control who sees your information, what’s shared, and what’s kept private.

Why does this matter for you? Because you’re active online and download things often, your personal info could be more exposed, making you a more attractive target for hackers or spam. If you don’t adjust these settings, strangers could see your posts, your email might get shared, or your browsing could be tracked and sold. This isn’t always obvious and can lead to scams or identity theft without you realizing it.

**Critical steps:**
- Set your profiles to private or “friends only.”
- Limit who can see your posts and personal info.
- Review which apps or services have access to your accounts and remove any you don’t use.
- Regularly check these settings—they can change over time.

Taking these simple steps is crucial to keep your information safe and to maintain your online privacy, which you clearly value. If you’re ever unsure, most services offer help guides—just search for “[service name] privacy settings.” 
\\
\hdashline

\textbf{Answer with conversation-based personalization:} \\
You can change your **Google privacy settings** in a few main places (these cover most of what people mean by “privacy settings”):

1) **Google Privacy Checkup (best starting point)**
- Go to: https://myaccount.google.com/privacycheckup  
- This walks you through key choices like what activity is saved, who can see your info, and ad personalization.

2) **Activity Controls (what Google saves about you)**
- Go to: https://myaccount.google.com/activitycontrols  
- Here you can turn on/off (or limit) things like **Web \& App Activity**, **Location History**, and **YouTube History**, and set **auto-delete**.

3) **My Activity (view/delete what’s already saved)**
- Go to: https://myactivity.google.com/  
- You can delete specific items or large time ranges. You can also enable **extra verification** so others can’t easily view it if they access your device.

4) **Ad Settings (reduce targeted ads)**
- Go to: https://adssettings.google.com/  
- Turn off **Ad personalization** to reduce profiling for advertising.

5) **Google Account “Data \& privacy” tab**
- Go to: https://myaccount.google.com/ → **Data \& privacy**  
- This is the “home” area where most privacy options live.

**Why this matters / risk if you ignore it:**  
If you never review these settings, Google may keep a long history of searches, locations, and YouTube activity. That can increase unwanted tracking and can be sensitive if someone else gets into your account or uses your signed-in device. This is **important but not an emergency**—it’s worth spending 10–15 minutes to review. \\

\\
\hline
\caption{Example questions and responses across personalization conditions.}
\label{tab:example_q_and_as}\\
\end{longtable}

\clearpage
\subsection{Evening and Final Survey Questions}
\label{sec:evening_final_survey_qs_appendix}

For question reviews, the evening and final surveys asked the same set of questions. Note a few small differences in the daily evening surveys and the final survey: 

First, participants were not asked to provide text why they did or did not follow the advice they received (this question was deemed too repetitive as participants already provided this information in the evening survey they filled out when they originally got the answer). 

Second, daily evening surveys asked for helpfulness rankings on a scale of 1-5 to provide more context for later personalization while the final survey asked participants to rank the helpfulness of answers on a scale of 1-4. The 1–4 scale was intentionally used in the final survey to remove a neutral midpoint and encourage participants to make a clearer comparative judgment between answers. Because the primary purpose of the final survey was to directly compare conversation-based personalization to the original answer received, we selected a forced-choice scale to better capture directional preferences rather than general satisfaction.

\begin{table}[H]
\centering
\small
\renewcommand{\arraystretch}{1.2} 
\begin{tabular}{|p{0.17\linewidth}@{\hspace{0.2cm}}|p{0.77\linewidth}@{\hspace{0.2cm}}|}
\hline

\multicolumn{2}{|l|}{\textbf{Evening Survey Questions About Answers Received During Day}} \\
\hline

\textbf{Understandability} & How many terms were there in the answer that were confusing or difficult to understand (1 = no terms, 3 = a couple, 5 = enough confusing terms that the answer was not understandable) \\
\hline

\textbf{Understandability} & Overall, how well were you able to understand the answer? (1 = not at all, 5 = I completely understood the answer) \\
\hline

\textbf{Actionability} & How easy was it to follow the advice given (if instructions were given)? ((1 = impossible to follow, 5 = very easy to follow) or n/a = no instructions given)\\
\hline

\textbf{Motivating Power} & Did you follow the advice provided by the assistant? (No, Somewhat, Yes) \\
\hline

\textbf{Motivating Power} & *Why or why not? (open text) \\
\hline

\textbf{Motivating Power} & Do you plan to follow the advice in the future? (Definitely yes, Probably yes, Might or might not, Probably not, Definitely not, NA (Already followed advice / no advice given)) \\
\hline

\textbf{General Helpfulness} & *On a scale of 1 - 5 how would you rate the answer you received to this question? (1 = not helpful at all, 5 = very helpful)\\
\hline

\multicolumn{2}{|l|}{\textbf{\color{blue}{Evening Survey Questions About Alternate Answers}}} \\
\hline

\textcolor{blue} {\textbf{General Helpfulness} }& \textcolor{blue}{*On a scale of 1 - 5 how would you rate the above alternate answer? (1 = not helpful at all, 5 = very helpful) }  \\
\hline

\textcolor{blue}{ \textbf{General Helpfulness} }& \textcolor{blue}{If you had received the above alternate answer would you have found it more or less helpful? (Much less helpful, Somewhat less helpful, About the same, More helpful, Much more helpful) }\\
\hline

\textcolor{blue}{ \textbf{Actionability} } & \textcolor{blue}{ If you had received the above alternate answer would you have been more or less likely to follow the advice provided (if applicable)? (Much less likely, Somewhat less likely, About the same level of likelihood, Somewhat more likely, Much more likely) }\\
\hline

\end{tabular}
\caption{Questions asked on the evening surveys and which metric these questions were addressing (if applicable). Questions pertaining to the alternate answers participants were shown in evening studies are shown in blue. Note that starred question differed slightly between daily evening surveys and final surveys. Why or why not? was not asked again in the final surveys, and helpfulness rankings were 1-4 in the final survey.}
\label{tab:evening_survey_questions}
\end{table}

\clearpage
\subsection{Relevance of User Questions to Real-World Context}
\label{sec:rel_to_real_world_appendix}

\begin{table}[H]
\centering
\small
\footnotesize
\setlength{\tabcolsep}{6pt}
\renewcommand{\arraystretch}{1.1}
\begin{tabular}{|l|c|c|}
\hline
\textbf{Option} & \textbf{Real Issue (\%)} & \textbf{Study Req. (\%)} \\ \hline
Almost never          & 5.26  & 21.05 \\
Rarely                & 5.26  & 28.07 \\
About half the time   & 31.58 & 29.82 \\
More often than not   & 33.33 & 17.54 \\
Almost all of the time & 24.56 & 3.51  \\ \hline
\textbf{weighted avg.} & \textbf{66.67} & \textbf{38.60} \\
\hline
\end{tabular}
\caption{Responses for question relevance. "Real Issue" are responses to the question "How many of the questions you asked during the study were related to real security issues or situations you encountered?"; "Study Req." are responses to the question "How often did you ask a question primarily to meet the study requirement, rather than because you were already wondering about it?". The weighted average maps each response category to an approximate proportion (Almost never = 0, Rarely = 0.25, About half = 0.5, More often than not = 0.75, Almost all of the time = 1) to estimate the overall fraction of questions in each category.}
\label{tab:survey-genuineness}
\end{table}

\subsection{Distribution of Question Times}
\label{sec:question_time_dist_appendix}

\clearpage
\subsection{Distribution of Question Topics}
\label{sec:q_topic_dist_appendix}

\begin{table}[H]
\centering
\small
\setlength{\tabcolsep}{4pt}  
\begin{tabular}{lcc}
\hline
\textbf{Topic} & \textbf{Exp.} & \textbf{Ctrl.} \\
\hline
Device Security & 62 & 31 \\
Password Security & 63 & 40 \\
Data Privacy & 46 & 46 \\
Secure Browsing Practices & 47 & 38 \\
Phishing Awareness & 46 & 56 \\
Network Security & 19 & 17 \\
Public Wi-Fi Risks & 31 & 29 \\
Identity Theft Prevention & 29 & 16 \\
Other & 27 & 30 \\
Two-Factor Authentication (2FA) & 22 & 19 \\
Malware and Viruses & 66 & 49 \\
User Behavior and Awareness & 31 & 37 \\
Cybersecurity Education & 28 & 14 \\
Security Tools and Software & 7 & 9 \\
Email Security & 11 & 20 \\
Incident Response & 3 & 6 \\
Software Updates & 12 & 6 \\
Social Engineering & 3 & 1 \\
Social Media Security & 4 & 11 \\
Children and Online Safety & 3 & 2 \\
Backup Strategies & 2 & 3 \\
Cloud Security & 5 & 6 \\
Remote Work Security & 0 & 3 \\
Legal and Ethical Considerations & 0 & 0 \\
Types of Cyberattacks & 4 & 4 \\
Emerging Technologies & 1 & 1 \\
\hline
\end{tabular}
\caption{Total number of questions falling under each topic with some questions falling under multiple topics, using the same topics as \citet{duesterwald2025}. The table separates topic numbers for questions under the experimental condition and questions asked by people in the control condition. As can be seen questions covered a wide range of topics in both conditions.}
\label{tab:question_topics}
\end{table}

\subsection{Sustained Interest in the Tool.}
\label{sec:sustained_intrest_appendix}

\begin{table}[h]
\centering
\small
\footnotesize
\setlength{\tabcolsep}{10pt}
\renewcommand{\arraystretch}{1.1}
\begin{tabular}{|l|c|}
\hline
\textbf{Option} & \textbf{Continue (\%)} \\ \hline
Yes   & 47.37 \\
Maybe & 45.61 \\
No    & 7.02  \\ \hline
\end{tabular}
\caption{Responses for continued access question. "Continue" is responses to the question "Would you be interested in continuing to have access to this tool?"}
\label{tab:survey-continue}
\end{table}

\clearpage
\subsection{Evening Survey Results}
\label{sec:evening_survey_res_appendix}

\begin{table*}[h]
\centering
\small
\renewcommand{\arraystretch}{1.2} 
\begin{tabular}{|
p{0.58\linewidth}@{\hspace{0.2cm}}|
p{0.125\linewidth}@{\hspace{0.2cm}}|
p{0.08\linewidth}@{\hspace{0.2cm}}|
p{0.085\linewidth}|}
\hline
\textbf{Survey Question} & \textbf{Experimental} & \textbf{Control} & \textbf{p-value} \\
\hline
\cellcolor{lightgreen} How many terms were there in the answer that were confusing or difficult to understand (1 = no terms, 3 = a couple, 5 = enough confusing terms that the answer was not understandable) & \cellcolor{lightgreen} 1.29545 & \cellcolor{lightgreen} 1.65789  & \cellcolor{lightgreen} 3.70E-07 \\
\hline
\textbf{Overall, how well were you able to understand the answer? (1 = not at all, 5 = I completely understood the answer)} & 4.73776 & 4.72672 & 0.80356 \\
\hline
\textbf{How easy was it to follow the advice given (if instructions were given)? (1 = impossible to follow, 5 = very easy to follow) (note: NA (no instructions were given) is excluded)} & 4.55769 & 4.62348 & 0.2095 \\
\hline
\textbf{Did you follow the advice provided by the assistant? (Yes=1, Somewhat=0, No=-1)} & 0.27622 & 0.33603 & 0.2439 \\
\hline
\cellcolor{lightgreen} Fraction of participants who answered ' Yes' that they followed the advice provided & \cellcolor{lightgreen} 0.51224 & \cellcolor{lightgreen} 0.583  & \cellcolor{lightgreen} 0.02434 \\
\hline
\textbf{Do you plan to follow the advice in the future? (Definitely yes = 2, Probably yes = 1, Might or might not = 0, Probably not = -1, Definitely not = -2)} & 1.44301 & 1.50429 & 0.25228 \\
\hline
Fraction of Participants who indicated either 'Yes' or 'Probably Yes' or 'Definitely Yes'  & 0.87413 & 0.90081 & 0.20311 \\
\hline
\cellcolor{lightgreen} Fraction of Participants who indicated either 'Yes' or “Definitely Yes' & \cellcolor{lightgreen} 0.73077 & \cellcolor{lightgreen} 0.80567  & \cellcolor{lightgreen} 0.00502 \\
\hline
\cellcolor{lightgreen} \textbf{On a scale of 1 - 5 how would you rate the answer you received to this question? (1 = not helpful at all, 5 = very helpful)} & \cellcolor{lightgreen} 4.43007 & \cellcolor{lightgreen} 4.70445  & \cellcolor{lightgreen} 8.91E-10 \\
\hline
\cellcolor{lightgreen} Fraction of answers that are helpful (4 or 5 / 5) & \cellcolor{lightgreen} 0.87587 & \cellcolor{lightgreen} 0.94939  & \cellcolor{lightgreen} 4.62E-05 \\
\hline

\multicolumn{4}{|l|}{\textbf{Questions Pertaining to Alternate Answers:}} \\
\hline
\cellcolor{lightgreen} \textbf{On a scale of 1 - 5 how would you rate the alternate answer you received to this question? (1 = not helpful at all, 5 = very helpful)} & \cellcolor{lightgreen} 4.34615 & \cellcolor{lightgreen} 4.51619 & \cellcolor{lightgreen} 0.00106 \\
\hline
\cellcolor{lightgreen} Fraction of alt. answers that are helpful (4 or 5 / 5) &  \cellcolor{lightgreen} 0.8549 &  \cellcolor{lightgreen} 0.90688 & \cellcolor{lightgreen} 0.01238 \\
\hline
\textbf{If you had received the above alternate answer would you have found it more or less helpful?} \textit{\color{blue}{(Much less helpful = -2, Somewhat less helpful = -1, About the same = 0, More helpful = 1, Much more helpful = 2) }} & 0.12413 & 0.10121 & 0.56514 \\
\hline
\textbf{If you had received the above alternate answer would you have been more or less likely to follow the advice provided (if applicable)?} \textit{\color{blue}{(Much less likely = -2, Somewhat less likely = -1, About the same level of likelihood = 0, Somewhat more likely = 1, Much more likely = 2)}} & 0.12903 & 0.13402 & 0.89184 \\
\hline
\cellcolor{lightgreen} Base answer helpful rating - alt. answer helpful rating & \cellcolor{lightgreen} 0.08392 & \cellcolor{lightgreen} 0.18826  & \cellcolor{lightgreen} 0.02457 \\
\hline
\end{tabular}
\caption{Results of the evening survey questions. Results for the experimental group (personalization condition where alternate answers had no personalization) and control group (no personalization condition where alternate answers had personalization) are averaged across all questions and all days. The leftmost column displays questions from the evening surveys (bolded) and statistics derived from answers directly provided to participants (not bolded). For questions with categorical values, the choices were mapped to numbers before means were calculated. These mappings are shown in blue and in italics. Rows highlighted in green are questions for which there was a significant difference between the results from the control and experimental groups (at a significance level of $\alpha$ = 0.05). Results come from 97 total participants (49 in experimental group, 48 in control) and a total of 1066 questions (572 coming from participants in the experimental group and 494 from the control)}
\label{tab:evening_survey_results_alt}
\end{table*}

\clearpage
\subsection{Directional Consistency Across LLM-Based and Human Evaluations}
\label{sec:cross_method_consistency_appendix}

\begin{table}[H]
\centering
\small
\renewcommand{\arraystretch}{1.2}
\setlength{\tabcolsep}{3.5pt}

\begin{tabular}{
|p{0.11\textwidth}|
p{0.13\textwidth}|
p{0.13\textwidth}|
p{0.13\textwidth}|
p{0.13\textwidth}|
p{0.13\textwidth}|
p{0.1\textwidth}|}
\hline
\textbf{Contrast and scale}
&
\multicolumn{2}{c|}{\textbf{Human evaluation}}
&
\multicolumn{3}{c|}{\textbf{LLM-based evaluation}}
&
\textbf{Direction agrees}
\\
\cline{2-6}

&
\textbf{Relative helpfulness}
&
\textbf{Relative likelihood of following advice}
&
\textbf{Understand- ability}
&
\textbf{Actionability}
&
\textbf{Motivating power}
&
\\
\hline

P4 $-$ P1     $[-2,2]$
&
$0.349 \pm 0.772$
&
$0.260 \pm 0.684$
&
$0.883 \pm 0.772$
&
$0.864 \pm 0.834$
&
$1.353 \pm 0.841$
&
Yes
\\
\hline

P4 $-$ PMT baseline $[-2,2]$
&
$0.145 \pm 0.522$
&
$0.164 \pm 0.460$
&
$0.547 \pm 0.582$
&
$0.478 \pm 0.609$
&
$0.709 \pm 0.610$
&
Yes
\\
\hline

\end{tabular}

\caption{Directional consistency between the human and LLM-based evaluations for the two contrasts represented in both settings. Cells report mean $\pm$ standard deviation on a $-2$ to $2$ comparative scale, where positive values favor P4. For P4 versus P1, the human evaluation included 175 relative-helpfulness evaluations and 169 relative-likelihood-of-following evaluations; the LLM-based evaluation included 531-532 matched participant-question evaluations, depending on the metric. For P4 versus the PMT-only baseline, the human evaluation included 110 evaluations for each outcome, and the LLM-based evaluation included 531-532 evaluations. LLM-based P4-versus-P1 values were computed by subtracting the P1-versus-baseline score from the P4-versus-baseline score for each matched participant-question pair.}
\label{tab:cross_method_directional_consistency}
\end{table}

\end{document}